\documentclass[
 reprint,
 amsmath,amssymb,
 aps,
 prd,
]{revtex4-1}

\usepackage{graphicx}
\usepackage{dcolumn}
\usepackage{bm}
\usepackage{hyperref}
\usepackage{color}
\usepackage{ulem}
\usepackage{xcolor}
\usepackage{textcomp}
\usepackage{etoolbox} 
\usepackage{lineno}
\apptocmd{\sloppy}{\hbadness 10000\relax}{}{} 
\definecolor{Nick}{RGB}{120, 94, 240}
\definecolor{Slaw}{RGB}{255, 165, 0}

\begin{document}

\title{Optical and mechanical properties of ion-beam-sputtered \texorpdfstring{Nb$_2$O$_5$}{Nb2O5} and \texorpdfstring{TiO$_2$-Nb$_2$O$_5$}{TiO2-Nb2O5} thin films for gravitational-wave interferometers}

\author{A. Amato, G. Cagnoli}
\affiliation{Universit\'{e} de Lyon, Universit\'{e} Claude Bernard Lyon 1, CNRS, Institut Lumi\`{e}re Mati\`{e}re, F-69622 Villeurbanne, France}

\author{M. Granata}
	\email[]{m.granata@lma.in2p3.fr}
\author{B. Sassolas}
\author{J. Degallaix}
\author{D. Forest}
\author{C. Michel}
\author{L. Pinard}
\affiliation{Laboratoire des Mat\'{e}riaux Avanc\'{e}s, Institut de Physique des 2 Infinis de Lyon, CNRS/IN2P3, Universit\'{e} de Lyon, Universit\'{e} Claude Bernard Lyon 1, F-69622 Villeurbanne, France}

\author{N. Demos}
	\email[]{ndemos@mit.edu}
\author{S. Gras}
\author{M. Evans}
\affiliation{Massachusetts Institute of Technology, 185 Albany Street NW22-295, Massachusetts 02139, Cambridge, USA}

\author{A. Di Michele}
\affiliation{Universit\`{a} degli Studi di Perugia, Dipartimento di Fisica e Geologia, Via Pascoli, 06123 Perugia, Italy}

\author{M. Canepa}
\affiliation{OPTMATLAB, Dipartimento di Fisica, Universit\`{a} di Genova, Via Dodecaneso 33, 16146 Genova, Italy}
\affiliation{INFN, Sezione di Genova, Via Dodecaneso 33, 16146 Genova, Italy}

\date{\today}

\begin{abstract}
Brownian thermal noise associated with highly-reflective mirror coatings is a fundamental limit for several precision experiments, including gravitational-wave detectors. Recently, there has been a worldwide effort to find mirror coatings with improved thermal noise properties that also fulfill strict optical requirements such as low absorption and scatter. We report on the optical and mechanical properties of ion-beam-sputtered niobia and titania-niobia thin films, and we discuss application of such coatings in current and future gravitational-wave detectors. We also report an updated direct coating thermal noise measurement of the HR coatings used in Advanced LIGO and Advanced Virgo.
\end{abstract}

\pacs{Valid PACS appear here}
\maketitle

Energy dissipation in amorphous coatings is a fundamental limitation for precision experiments such as interferometric gravitational-wave detectors (GWDs) \cite{Adhikari14}, optomechanical resonators \cite{Aspelmeyer14}, frequency standards \cite{Matei17}, and quantum supercomputers \cite{Martinis05}. In these devices, thermally-driven random structural relaxations distribute the thermal energy of the normal modes of vibration across a wide frequency range, giving rise to Brownian {\it coating thermal noise} (CTN) \cite{Saulson90,Levin98}. The power spectral density of such thermally induced surface fluctuations is determined by the rate of energy dissipation in the coating material, as stated by the fluctuation-dissipation theorem \cite{Callen52}. The CTN power spectral density $S_{\textrm{CTN}}$, as measured with an optical beam, can be written in the simplified form \cite{Harry02}
\begin{equation}\label{eqn.S}
S_{\textrm{CTN}} \propto \frac{k_B T}{2\pi f} \frac{d}{\omega^2} \varphi_c(f)\ ,
\end{equation}
where $k_B$ is the Boltzmann constant, $f$ is the frequency, $T$ is the temperature, $d$ is the coating thickness, $\omega$ is the laser beam radius where intensity drops by 1/e$^2$, and $\varphi_c(f)$ is the loss angle associated with energy dissipation in the coating. The loss angle quantifies the internal mechanical friction in the coating material and is defined as the ratio of the imaginary to real parts of the elastic modulus, $\varphi_c(f)\equiv \text{tan}^{-1}[\text{Im}{(Y_c)}/\text{Re}{(Y_c)}]$. 

Thermally induced surface fluctuations can be reduced by increasing the beam radius $\omega$, decreasing the temperature $T$, or by choosing coating materials which minimize $d\varphi_c$. Increases in $\omega$ are limited by both the difficulties in uniformly coating larger substrates and by arm cavity control stability, while decreases in $T$, such as operating in the cryogenic regime, are limited by experimental complexity issues \cite{KAGRA,ET,Abbott17} and by the narrow selection of materials which are known to have favorable properties at cryogenic temperatures.

High-reflection (HR) optical coatings are usually Bragg reflectors of alternating layers of low- and high-refractive-index materials, where the number of low/high index pairs determines the coating transmissivity. However, for the same transmissivity, the number of pairs can vary depending on the refractive index contrast $C = n_{\textrm{\tiny{H}}}/n_{\textrm{\tiny{L}}}$, where $n_{\textrm{\tiny{H}}}$ and $n_{\textrm{\tiny{L}}}$ are the high and low refractive indices, respectively: the higher the $C$, the lower the coating thickness $d$ and hence the CTN. The high index material is usually the most dissipative one \cite{Granata20,Penn03,Crooks04}. Ideally, for a given $n_{\textrm{\tiny{L}}}$, the high index material should have the highest $n_{\textrm{\tiny{H}}}$ possible in order to maximize $C$ and reduce its physical thickness.

The HR coatings of the Advanced LIGO \cite{aLIGO}, Advanced Virgo \cite{AdVirgo} and KAGRA \cite{KAGRA} GWDs are thickness-optimized stacks \cite{Villar10} of ion-beam-sputtered (IBS) layers of tantalum pentoxide (Ta$_2$O$_5$, also known as {\it tantala}, high index) and silicon dioxide (SiO$_2$, {\it silica}, low index), produced by the Laboratoire des Mat\'{e}riaux Avanc\'{e}s (LMA) \cite{Degallaix19,Pinard17}. Following a procedure developed by the LMA \cite{Comtet07} for the LIGO Scientific Collaboration \cite{Harry07} in order to reduce their optical absorption and lower their loss angle, the high-index layers of Advanced LIGO and Advanced Virgo are a uniform mixture of co-sputtered tantala and titanium dioxide (TiO$_2$, {\it titania}) \cite{Granata20}.

Despite their superb optical and mechanical properties \cite{Degallaix19,Granata20}, the CTN of current HR coatings remains a severe limitation for further sensitivity improvement in GWDs. In the last two decades, a considerable research effort has been committed to finding an alternative high-index material featuring both low mechanical loss and low optical loss (absorption, scattering) \cite{Granata20review}.

In this paper we report on the optical and mechanical properties of IBS niobium pentoxide (Nb$_2$O$_5$, {\it niobia}) and titania-niobia (TiO$_2$-Nb$_2$O$_5$) thin films, and we discuss application of niobia-based coatings in current and future GWDs. As we wanted to compare the CTN of those newly-developed coatings against that of current HR coatings of GWDs, we also improved our direct CTN measurement method and are able to report an updated CTN of the HR coatings used in Advanced LIGO and Advanced Virgo.

\section{Methods}
Single thin layers (470 to 490 nm thick) of IBS Nb$_2$O$_5$ and TiO$_2$-Nb$_2$O$_5$ coatings were deposited on silicon wafers ($\varnothing$ 2") to measure their optical properties and on fused-silica disk-shaped resonators ($\varnothing$ 50 mm, 1 mm thick) to measure their mechanical properties. In order to fully test an alternative design for current GWDs, HR stacks for operation at 1064 nm were deposited on fused-silica witness samples ($\varnothing$ 1") for optical and CTN measurements.

All coatings were deposited in a custom-developed IBS coating machine, using accelerated, neutralized argon ions as sputtering particles. Prior to deposition, the base pressure inside the coater vacuum chamber was less than $10^{-7}$ mbar. Argon (12 sccm) was fed into the ion-beam source while oxygen (20 sccm) was fed into the chamber, for a total pressure of the order of $10^{-4}$ mbar inside the chamber. Energy and current of the sputtering Ar ions were 1.0 keV and 0.2 A, respectively. The source-target and target-substrate angles were set to 45$^{\circ}$. During deposition, the sputtered coating particles (co-sputtered, in the case of titania-niobia coatings) impinged on substrates heated up to about 100 $^{\circ}$C.

After deposition, the coated samples were annealed in air for 10 hours at $T_a = 400$ $^\circ$C. Annealing is a standard procedure to decrease the internal stress, the optical absorption and the internal friction of coatings \cite{Granata20, Amato19b}. The maximum annealing temperature $T_a$ is limited by the onset of crystallization, which makes the amorphous coatings undergo a phase change and become poly-crystalline.
 
\subsection{Optical characterization}
\begin{figure*}
\centering
	\includegraphics[width=0.75\textwidth]{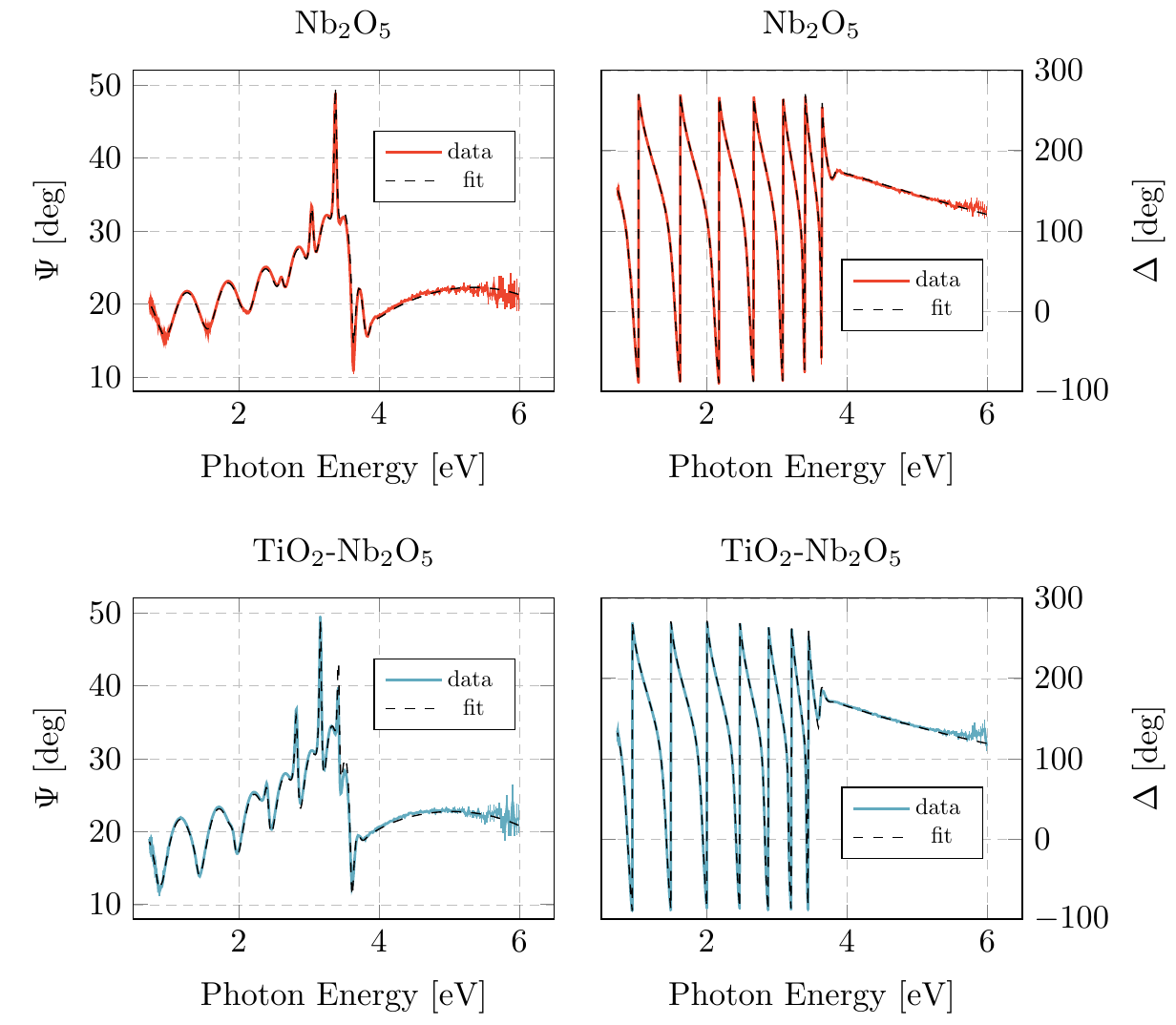}
	\caption{Ellipsometric spectra of annealed Nb$_2$O$_5$ (top row) and TiO$_2$-Nb$_2$O$_5$ (bottom row) films, acquired at an incidence angle $\theta$ = 60$^\circ$.}
\label{FIGspectra}
\end{figure*}

We used two J. A. Woollam spectroscopic ellipsometers to measure optical properties and thickness of the single-layer coatings, covering complementary spectral regions: a VASE for the 190-1100 nm range and a M-2000 for the 245-1680 nm range. The wide range swept with both instruments allowed us to extend the analysis from ultraviolet to near infrared (0.7 - 6.5 eV). The coated silicon wafers (with only one-side polished to suppress spurious reflections from the rear surface) were measured in reflection. The optical properties were obtained by measuring the amplitude ratio $\Psi$ and phase difference $\Delta$ of the p- and s-polarized reflected light \cite{Fujiwara07}. To maximize the response of the instruments, the ($\Psi$, $\Delta$) spectra were acquired for three different incidence angles ($\theta$ = 50$^\circ$, 55$^\circ$, 60$^\circ$), chosen to be close to the Brewster angle of the coatings. Coating refractive index and thickness were derived by fitting the experimental data with the well-known Cody-Lorentz \cite{Ferlauto02} and Tauc-Lorentz \cite{Jellison96} optical models, the optical response of the bare wafers were characterized with prior dedicated measurements. By way of example, Fig. \ref{FIGspectra} shows the ($\Psi$, $\Delta$) spectra of the annealed niobia and titania-niobia coatings. Fig. \ref{FIGrefrInd} shows the reconstructed dispersion laws. Further details about our ellipsometric analysis are available elsewhere \cite{Amato19,Amato20}.

To measure scattering and optical absorption of the HR stacks at $\lambda$ = 1064 nm (the operational wavelength of current GWDs), we characterized the coated fused-silica witness samples with a commercial CASI scatterometer and a custom-developed setup \cite{Beauville04} based on the photo-thermal deflection principle \cite{Boccara80}, respectively.

\subsection{Mechanical characterization}
Before and after each treatment (coating deposition, annealing), we measured the mass of the disks with an analytical balance and used the measured coating thickness values to calculate the coating density $\rho_c$.

We used the ring-down method \cite{Nowick72} to measure the frequency $f$ and ring-down time $\tau$ of the first vibrational modes of each disk, before and after the coating deposition, and calculated the coating loss angle
\begin{equation}
\label{EQcoatLoss}
\varphi_c = \frac{\varphi + (D-1)\varphi_0}{D} \ ,
\end{equation}
where $\varphi_0 = (\pi f_0 \tau_0)^{-1}$ is the measured loss angle of the bare substrate, $\varphi = (\pi f \tau)^{-1}$ is the measured loss angle of the coated disk, and $D$ is the frequency-dependent measured \textit{dilution factor} \cite{Li14}. We measured up to eight modes, from $\sim$2.5 to $\sim$33 kHz, in a frequency band which partially overlaps with the detection band of ground-based GWDs (10-10$^4$ Hz). In order to avoid systematic damping from suspension and residual gas pressure, we used a clamp-free in-vacuum Gentle Nodal Suspension (GeNS) system \cite{Cesarini09}. This system is currently the preferred solution of the Virgo and LIGO Collaborations for performing internal friction measurements \cite{Granata20,Vajente17}.

The coating Young modulus $Y_c$ and Poisson ratio $\nu_c$ were estimated by fitting finite-element simulations to measured dilution factors via least-squares numerical regression \cite{Granata20}. Fig. \ref{FIGdilFact} shows the results of this analysis. Further details about our GeNS system, finite-element simulations and data analysis are available elsewhere \cite{Granata20,Granata16}.

\subsection{Thermal noise}\label{sec.th}
The direct CTN measurements were conducted with a multi-mode technique \cite{Gras18,Gras17} using a folded standing-wave Fabry-Perot cavity, where the folding mirror was the coating sample. Each of the co-resonating second-order orthogonal transverse modes, TEM02 and TEM20, act as a cavity length sensor for a different region of the coating surface. Any noise common to both TEMs cancels out on the beat frequency between these modes, whereas noise from thermally induced vibrations in separate regions of the coating adds in quadrature. The beat frequency, usually $\sim 5$ MHz, occurs as a result of astigmatism in the cavity end mirrors. This technique allows for the routine measurement of thermal noise spectra in the frequency band from $\sim$30 Hz to $\sim$3 kHz. The large sensitivity of this experimental setup also allows us to measure the cavity response to thermal load and thus estimate the coating optical absorption. The multi-mode technique is currently the preferred method of the LIGO Collaboration for performing direct CTN measurements of HR coatings.

In previous direct CTN measurements that used this technique \cite{Gras18,Gras17}, we assumed that the cavity end mirrors contributed negligibly to the measured CTN. For the results presented here, however, we properly accounted for their contribution. To measure the CTN contributions from the cavity end mirrors, we repeatedly measured a sample with various combinations of end mirrors. In addition, our newest set of end mirrors were coated along with a witness sample that we could then directly measure. Using these measurements we were able to fit the cavity end mirror CTN and subtract it from our data. We found that the CTN from the cavity end mirrors contributes a small but non-negligible amount to the total measured CTN. This measurement also allowed us to update our reported value of the CTN amplitude of the coatings currently in use by Advanced LIGO and Advanced Virgo.

\subsection{Composition}
We used a Zeiss LEO 1525 field-emission scanning electron microscope (SEM) and a Bruker Quantax system equipped with a Peltier-cooled XFlash 410-M silicon drift detector to analyze the surface and elemental composition of the as-deposited titania-niobia coatings. Semi-quantitative (standardless) results were based on a peak-to-background evaluation method of atomic number, self-absorption and fluorescence effects (P/B-ZAF correction) and a series fit deconvolution model provided by the Bruker Esprit 1.9 software. Using the self-calibrating P/B-ZAF standard-based analysis, no system calibration had to be performed.

The SEM beam was set to 15 keV for the surface survey. We performed multiple energy-dispersive X-ray (EDX) analyses on different coating sample spots and with different magnifications (from $100\times$ to $5000\times$), for a total scanned surface of $\sim$5 mm$^2$.

\section{Results}

\subsection{Single layers}
\begin{figure*}
\centering
	\includegraphics[width=0.75\textwidth]{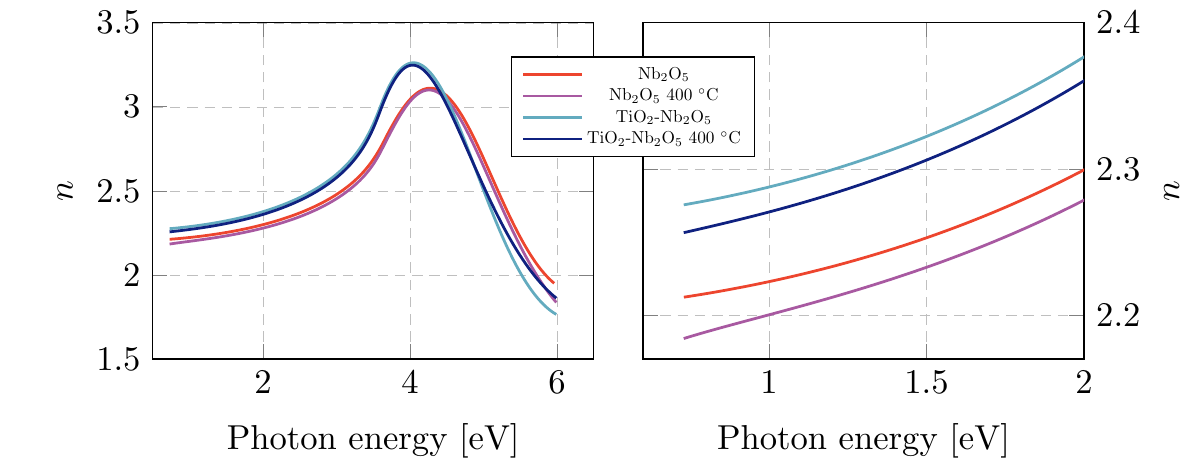}
	\caption{Refractive index $n$ of Nb$_2$O$_5$ and TiO$_2$-Nb$_2$O$_5$ thin films, before and after in-air annealing at 400 $^{\circ}$C for 10 hours. The right plot is a zoom on the region of interest for present and future GWDs ($\lambda = 1064$ nm corresponds to a photon energy $E$ of 1.17 eV, $\lambda = 1550$ nm to $E = 0.80$ eV).}
\label{FIGrefrInd}
\end{figure*}

\begin{figure}
\centering
	\includegraphics[width=0.45\textwidth]{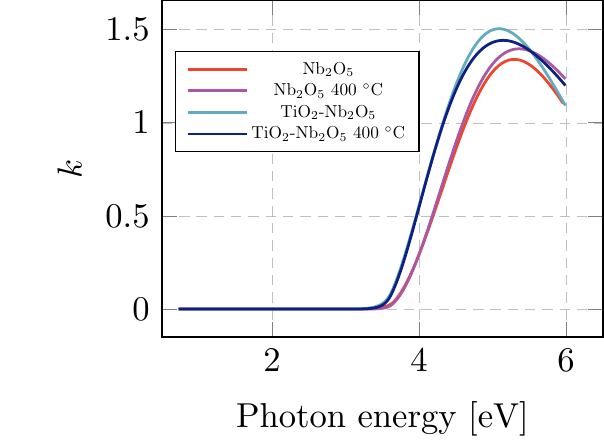}
	\caption{Extinction coefficient $k$ of Nb$_2$O$_5$ and TiO$_2$-Nb$_2$O$_5$ thin films, before and after in-air annealing at 400 $^{\circ}$C for 10 hours.}
\label{FIGextinctCoeff}
\end{figure}

Several niobia samples were analyzed by spectroscopic ellipsometry, all yielding consistent results, together with a single titania-niobia sample. Fig. \ref{FIGspectra} shows exemplary $(\Psi,\Delta)$ spectra for both sets of annealed samples, acquired at $\theta$ = 60$^\circ$. All spectra showed a degradation of the signal to noise ratio above 5.3 eV, caused by strong absorption. Above 6 eV, the signal to noise ratio was drastically reduced and data was discarded since no longer useful for fitting purposes. Our models fit all the measured spectra with the same accuracy. Figures \ref{FIGrefrInd} and \ref{FIGextinctCoeff} show the dispersion laws and the extinction coefficient derived from our analysis, respectively.

On Fig. \ref{FIGspectra}, the interference features (due to multiple reflections in the  transparency region) stop quite sharply at the fundamental absorption threshold. By approaching this threshold, in the visible region, the oscillation amplitude gradually decreases. Within the boundaries of the measurement uncertainty, the energy gap is the same for niobia and titania-niobia coatings, $E_g = 3.4 \pm 0.1$ eV. This can be explained by the fact that the energy gap of our IBS titania coatings is $E_g = 3.45 \pm 0.05$ eV, i.e. very similar to that of niobia coatings. However, the refractive index of titania-niobia coatings increased substantially in the infrared region with respect to that of niobia coatings. At 1064 nm, we found $n = 2.24 \pm 0.01$ for niobia and $n = 2.30 \pm 0.01$ for titania-niobia coatings, before annealing. Our results for niobia coatings are consistent with values found in the literature for IBS coatings deposited with various sputtering settings \cite{Lee00,Lee02,Cetinorgu09}. For comparison, the refractive index at 1064 nm of our IBS titania coatings is $n = 2.35 \pm 0.05$ before annealing \cite{Granata20}.

\begin{figure*}
\centering
	\includegraphics[width=0.75\textwidth]{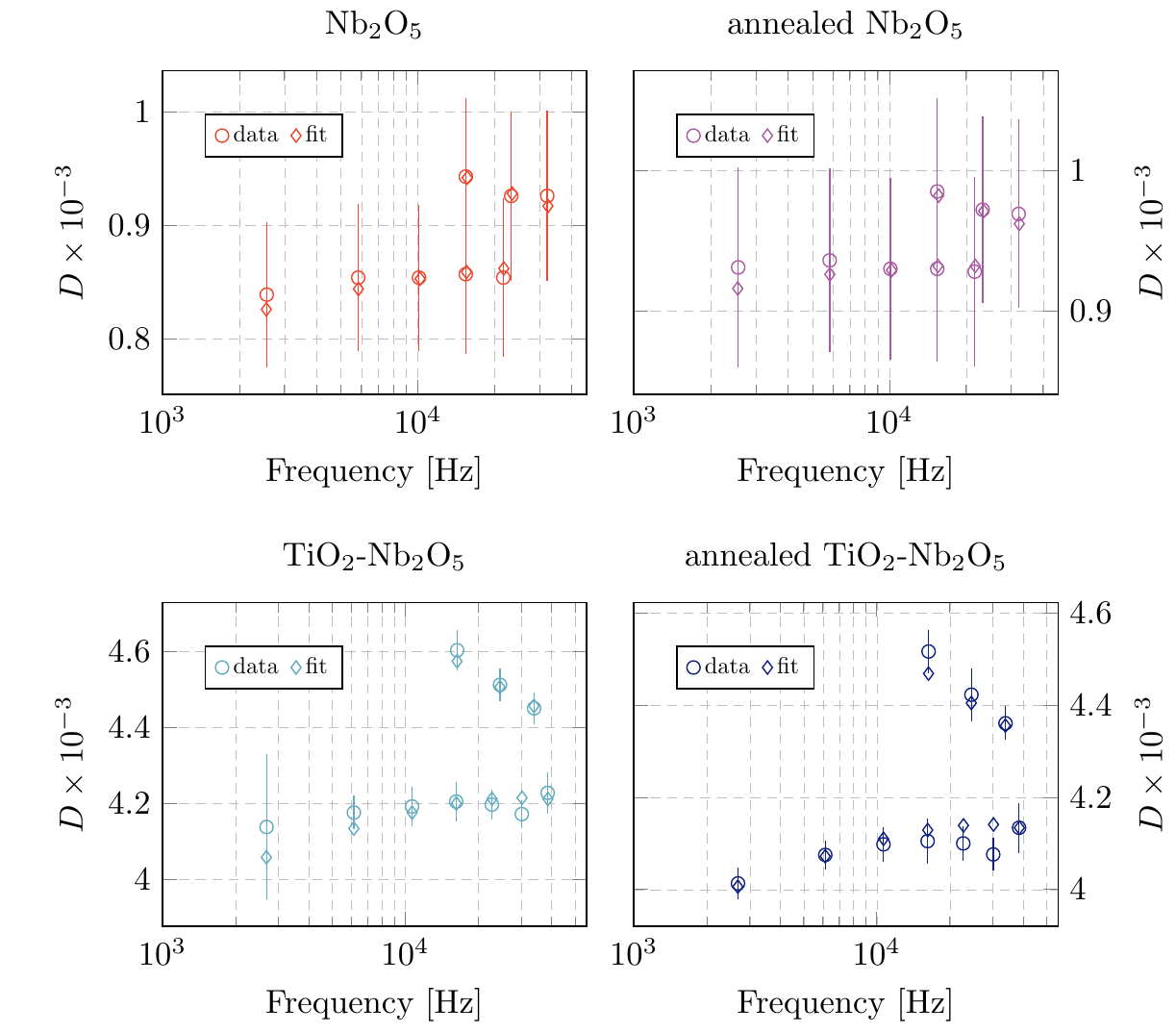}
	\caption{Dilution factor $D$ of Nb$_2$O$_5$ and TiO$_2$-Nb$_2$O$_5$ films, before and after in-air annealing at 400 $^{\circ}$C for 10 hours.}
\label{FIGdilFact}
\end{figure*}

\begin{figure*}
\centering
	\includegraphics[width=0.75\textwidth]{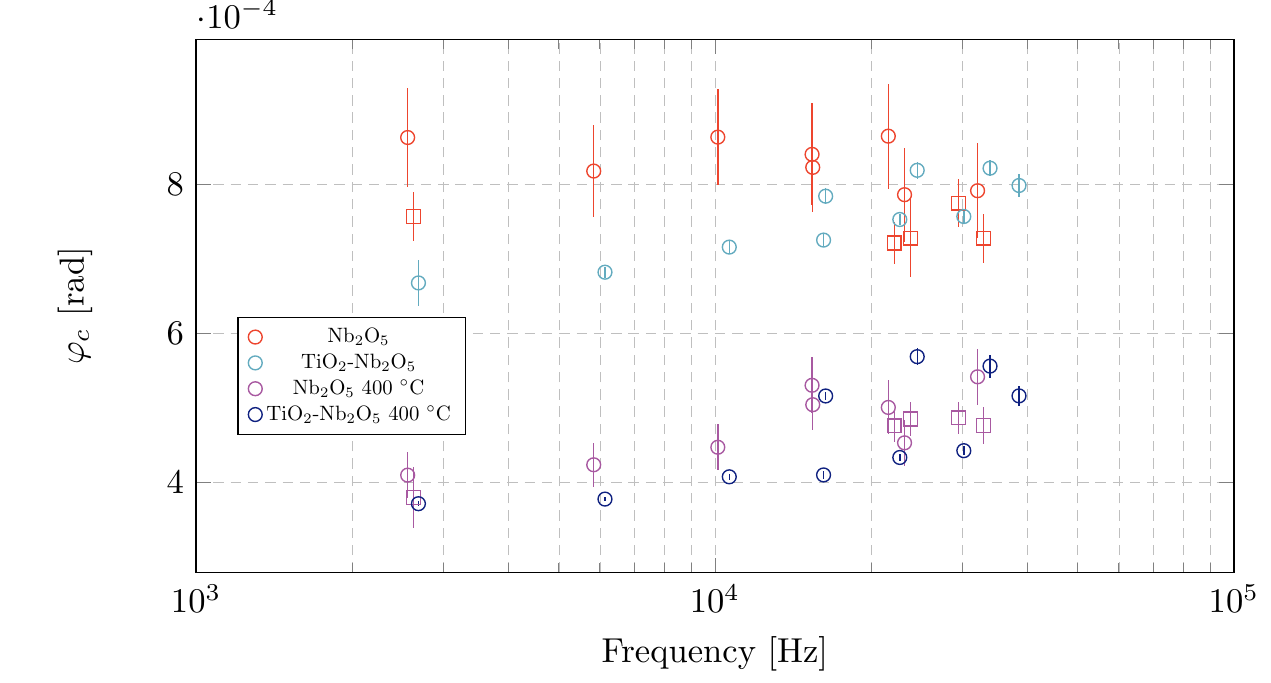}
	\caption{Mechanical loss $\varphi_c$ of Nb$_2$O$_5$ and TiO$_2$-Nb$_2$O$_5$ films, before and after in-air annealing at 400 $^{\circ}$C for 10 hours (different markers indicate distinct samples).}
\label{FIGcoatLoss}
\end{figure*}

For the mechanical properties, we characterized two disks coated with niobia films and a disk coated with titania-niobia films. Fig. \ref{FIGdilFact} shows measured dilution factors of a sample from each coating set. Our finite-element simulations fitted the data of all samples with similar accuracy. Fig \ref{FIGcoatLoss} shows the measured coating loss angles, calculated using Eq.(\ref{EQcoatLoss}).

For the as-deposited niobia films, we found $\rho_c = 4.4 \pm 0.7$ g/cm$^{3}$, $Y_c = 100 \pm 7$ GPa and $\nu_c = 0.30 \pm 0.02$. For comparison, values reported previously by \c{C}etin\"{o}rg\"{u} et al. \cite{Cetinorgu09} are $\rho_c = 4.50$ g/cm$^{3}$, a reduced Young modulus of $118$ GPa and $\nu_c = 0.22$. Concerning the elastic constants, our values are substantially different from what can be found in the literature. This discrepancy might be explained by the different sputtering settings used to produce the samples, as observed for other high-index oxide coatings \cite{Granata20}, and by the different methods used for the measurement. For the Young modulus in particular, \c{C}etin\"{o}rg\"{u} et al. used nanoindentation, which produces results that may vary depending on the substrate used for the coating deposition \cite{Granata20} and on the model used for the analysis. Unlike nanoindentation, our method yields consistent results for the same coating on different substrates \cite{Granata20} and does not rely on any specific assumption about the model to be used for data analysis and on the actual value of $\nu_c$.

According to our EDX analyses, the titania-niobia coatings feature an average atomic cation ratio Nb/(Ti+Nb) $=0.27 \pm 0.01$. Compared to niobia coatings, the co-sputtering induced a moderate decrease of the coating loss angle and significantly increased the coating Young modulus.

The maximum annealing temperature $T_a$, limited by the onset of crystallization, was 400 $^{\circ}$C for both niobia and titania-niobia coatings. For comparison, $300 < T_a < 350$ $^{\circ}$C for IBS titania \cite{Lee06,Chen08}. We observed that annealing increased the coating thickness by $2-3$\% and slightly reduced the refractive index by about $1$\% in the transparency region, whereas it did not change the energy gap. Further analysis of the ($\Psi$, $\Delta$) spectra also found for both niobia and titania-niobia coatings the same effect already observed in IBS tantala and tantala-titania coatings, that is, a reduction of the Urbach tails \cite{Amato19b}. Remarkably, annealing decreased the coating loss angle of a factor $1.8-2$ at $\sim$2.5 kHz. 

Table \ref{TABLEmonolayers} lists the measured optical and mechanical properties of our IBS niobia and titania-niobia coatings. Refractive index $n$ is given for the wavelength of operation of current GWDs, $\lambda=1064$ nm (corresponding to a photon energy $E$ of 1.17 eV), as well as for the alternative wavelength $\lambda=1550$ nm ($E = 0.80$ eV) of future GWDs such as Einstein Telescope \cite{ET}.
\begingroup
\squeezetable
\begin{table*}
\caption{\label{TABLEmonolayers} Optical and mechanical properties of IBS Nb$_2$O$_5$ and TiO$_2$-Nb$_2$O$_5$ coatings, before and after 400 $^{\circ}$C annealing: refractive index $n$ at 1064 and 1550 nm, energy gap $E_g$, density $\rho_c$, loss angle $\varphi_c$ at $\sim$2.5 kHz (from direct ring-down measurements and fit to CTN data), Young modulus $Y_c$ and Poisson ratio $\nu_c$. Loss angle extracted from CTN measurements assume low-index material loss angle 2.3$\times10^{-5}$ rad \cite{Granata20}. Values of Ta$_2$O$_5$-TiO$_2$ layers of Advanced LIGO and Advanced Virgo are also listed, for comparison \cite{Granata20,Amato19}.}
\begin{ruledtabular}
\begin{tabular}{lcccccccc}
	& $n_{\text{1064}}$	& $n_{\text{1550}}$	& $E_g$ [eV]	& $\rho_c$ [g/cm$^{3}$]	& $\varphi_c$ [$10^{-4}$ rad]	& $\varphi_c^{\textrm{\tiny{CTN}}}$ [$10^{-4}$ rad] & $Y_c$ [GPa] & $\nu_c$\\
	\colrule
Nb$_2$O$_5$ & 2.24 $\pm$ 0.01 & 2.22 $\pm$ 0.01 & 3.4 $\pm$ 0.1 & 4.4 $\pm$ 0.7 & 8.1 $\pm$ 0.5 & -- & 100 $\pm$ 8 & 0.30 $\pm$ 0.03\\
Nb$_2$O$_5$ 400 $^{\circ}$C & 2.22 $\pm$ 0.01 & 2.20 $\pm$ 0.01 & 3.4 $\pm$ 0.1 & 4.2 $\pm$ 0.7 & 3.9 $\pm$ 0.1 & 7.2 $\pm$ 0.6 & 99 $\pm$ 2 & 0.24 $\pm$ 0.02\\
TiO$_2$-Nb$_2$O$_5$ & 2.30 $\pm$ 0.01 & 2.28 $\pm$ 0.01 & 3.3 $\pm$ 0.1 & 4.3 $\pm$ 0.1 & 6.7 $\pm$ 0.3 & -- & 120 $\pm$ 1 & 0.30 $\pm$ 0.01\\
TiO$_2$-Nb$_2$O$_5$ 400 $^{\circ}$C & 2.28 $\pm$ 0.01 & 2.26 $\pm$ 0.01 & 3.3 $\pm$ 0.1 & 4.1 $\pm$ 0.1 & 3.7 $\pm$ 0.1 & 8.2 $\pm$ 1.1 / 10.2 $\pm$ 1.8 & 116 $\pm$ 1 & 0.29 $\pm$ 0.01\\
Ta$_2$O$_5$-TiO$_2$ 500 $^{\circ}$C & 2.09 $\pm$ 0.01 & 2.08 $\pm$ 0.01 & 3.6 $\pm$ 0.1 & 6.7 $\pm$ 0.1 & 3.4 $\pm$ 0.3 & 5.5 $\pm$ 0.9 & 120 $\pm$ 4 & 0.29 $\pm$ 0.01\\
\end{tabular}
\end{ruledtabular}
\end{table*}
\endgroup

\subsection{HR stacks}
We produced three different Bragg reflectors with the following materials and designs for operation at 1064 nm: (i) niobia and silica layers of quarter-wavelength ($\lambda/4$) optical thickness; (ii) titania-niobia and silica layers of $\lambda/4$ optical thickness; (iii) titania-niobia and silica layers of optimized thickness \cite{Villar10}, to minimize the thickness of the more dissipative high-index layers. Thus, if $d_{\textrm{\tiny{H}}}$ and $d_{\textrm{\tiny{L}}}$ are  the cumulative thicknesses of high- and low-index layers, respectively, the optimization allowed to reduce the thickness ratio $\xi \equiv d_{\textrm{\tiny{H}}}/d_{\textrm{\tiny{L}}}$ down to 38\% in sample (iii), compared to $\sim$60\% of samples (i) and (ii). 

All HR samples were designed to yield 5 parts per million (ppm) transmissivity, as in the current HR coatings of the end test masses (ETMs) of Advanced LIGO and Advanced Virgo \cite{Granata20}. Their design specifications are summarized in Table \ref{TABLE_HRstacks}, where they are also compared to current ETM coatings: all our alternative stacks are thinner, thanks to their higher index contrast $C$, and sample (iii) features the lowest ratio $\xi$ and hence the smallest content
of dissipative material. However, because of the  transmissivity requirement, the optimization of sample (iii) came with the cost of an increased thickness.

We characterized the optical properties of our HR samples after they were annealed at 400 $^{\circ}$C for 10 hours in air. For each sample, we measured 4.3 ppm transmission and 0.3 ppm absorption. This is the same value of absorption as measured in current ETMs and represents a factor 7 improvement over our previous niobia coatings \cite{Flaminio10}. Cavity response measurements (see Section \ref{sec.th}) give comparable results for sample (i) and a factor of 2 and 4 higher for samples (ii) and (iii), respectively. Sample (iii) was measured in two different spot locations, while samples (i) and (ii) were measured at only one spot. Table \ref{TABLE_CTNresults} lists the measured optical properties.
\begin{table*}
\caption{\label{TABLE_HRstacks} Nominal specifications (layers, design, number of layers $N$, thickness of high-index layers $d_H$, thickness of low-index layers $d_L$, thickness ratio $\xi = d_{\textrm{\tiny{H}}}/d_{\textrm{\tiny{L}}}$, total thickness $d = d_{\textrm{\tiny{H}}} + d_{\textrm{\tiny{L}}}$) and measured properties (optical absorption $\alpha$ from photo-thermal deflection and CTN measurements, scattering $\alpha_s$) of HR coatings for 5 ppm reflectivity. Coatings were measured after annealing (400 $^{\circ}$C for 10 hours, in air). Values of Ta$_2$O$_5$-TiO$_2$/SiO$_2$ ETM coatings of Advanced LIGO and Advanced Virgo \cite{Pinard17,Granata20,Gras18} are also listed, for comparison. CTN absorption values are relative to $\alpha = 0.27$ for the ETM coatings.}
\begin{ruledtabular}
\begin{tabular}{cccccccccc}
	& design & $N$ & $d_H$ [nm] & $d_L$ [nm]	& $\xi$	& $d$ [nm] & $\alpha$ [ppm] & $\alpha^{\textrm{\tiny{CTN}}}$ [ppm] & $\alpha_s$ [ppm]\\
	\colrule
	Nb$_2$O$_5$/SiO$_2$			& $\lambda/4$ & 33 & 1955 & 3136 & 0.62 & 5091 & 0.32 $\pm$ 0.05 & 0.23 & 6 $\pm$ 2\\
	TiO$_2$-Nb$_2$O$_5$/SiO$_2$	& $\lambda/4$ & 31 & 1774 & 2951 & 0.60 & 4726 & 0.35 $\pm$ 0.05 & 0.44 & 32 $\pm$ 1\\
	TiO$_2$-Nb$_2$O$_5$/SiO$_2$	& optimized & 34 & 1545 & 4090 & 0.38 & 5635 & 0.30 $\pm$ 0.05 & 0.99 & 25 $\pm$ 1\\
	ETM	& optimized & 38 & 2109 & 3766 & 0.56 & 5875 & 0.27 $\pm$ 0.07 & & 5 $\pm$ 2\\\end{tabular}
\end{ruledtabular}
\end{table*}

After annealing, bubble-like defects of different number and size, detected with an optical microscope, appeared at variable depth in the coatings. We observed no trace of such defects in the annealed single layers, indicating that this phenomenon only occurs when layers are stacked. These defects might be caused by clustering of incorporated argon atoms which, according to our EDX analyses, amounts to an atomic concentration of $0.7 \pm 0.1$ \% in the titania-niobia layers of samples (ii) and (iii). Samples (ii) and (iii) scattered 5 to 6 times more light than what is measured for current ETMs, which may be explained by the observation of numerous defects. By contrast, sample (i) featured a reduced number of defects and the same scattering value as the ETMs, 6 ppm. Further analyses will be needed to determine the size and spatial distribution statistics of defects, and to correlate it with sputtering settings, annealing parameters (temperature, duration, atmosphere) and measured scattering.
\begin{table*}
\caption{\label{TABLE_CTNresults} Results of direct CTN measurements at 100 Hz and 50 \textmu m beam size (amplitude, ratio with respect to ETM witness sample, power index of frequency dependence) and loss angle $\varphi_c$ of high-index material of HR coatings for 5 ppm transmission. Loss angle calculation assume low-index material loss angle 2.3$\times10^{-5}$ rad \cite{Granata20}. Coatings have been measured after annealing (400 $^{\circ}$C for 10 hours, in air). Values of Ta$_2$O$_5$-TiO$_2$/SiO$_2$ ETM coatings of Advanced LIGO and Advanced Virgo \cite{Pinard17,Granata20,Gras18} are also listed, for comparison.
}
\begin{ruledtabular}
\begin{tabular}{cccccc}
	& design & amplitude [10$^{-18}$ m/$\sqrt{\text{Hz}}$] & ratio to ETM	& slope	& $\varphi_c$ [10$^{-4}$ rad]\\
	\colrule
	Nb$_2$O$_5$/SiO$_2$	& $\lambda/4$ & 13.4 $\pm$ 0.1 & 1.04 & 0.45 $\pm$ 0.01 & 5.2 $\pm$ 0.1\\
	TiO$_2$-Nb$_2$O$_5$/SiO$_2$ & $\lambda/4$ & 12.7 $\pm$ 0.3 & 0.98 & 0.42 $\pm$ 0.01 & 4.9 $\pm$ 0.3\\
	TiO$_2$-Nb$_2$O$_5$/SiO$_2$ & optimized & 13.2 $\pm$ 0.2 & 1.02 & 0.42 $\pm$ 0.02 & 6.1 $\pm$ 0.2\\
	ETM & optimized	& 12.9 $\pm$ 0.2 & 1.00 & 0.45 $\pm$ 0.02 & 4.0 $\pm$ 0.1\\
\end{tabular}
\end{ruledtabular}
\end{table*}

Direct CTN measurements were taken at two different locations, separated by less than 200 $\mu$m, on each of the three samples. An amplitude spectral density (ASD) measurement was repeated at least three times in each location. For each sample, the variation in ASD between the two locations was less than 3\%, thus we treated these measurements as statistically identical and report their variance weighted mean values in Table \ref{TABLE_CTNresults}. Fig. \ref{FIGctn_asd} shows an exemplary ASD measurement of the stack containing niobia. Although the direct CTN measurement is more sensitive to individual defects, because the beam spot size used in the measurement is small, we found no evidence that our measurement was contaminated by the presence of bubble-like defects. Using these measurements we extracted the CTN amplitude ($N_{CTN}$) at 100 Hz as well as the CTN frequency dependence (slope),
\begin{equation}
    N_{CTN}(f) = \textrm{amplitude} \times \left( \frac{\text{100 Hz}}{f} \right)^{\textrm{slope}}
\end{equation}
using least-squares and Monte Carlo fitting.
We found that the CTN of sample (i) was 4\% larger, sample (ii) was 2\% lower, and sample (iii) was 2\% larger than the CTN of current ETMs. The CTN amplitude frequency dependence of sample (i) matched that of the ETM samples ($f^{-0.45\pm0.02}$), while the frequency dependence of samples (ii) and (iii) was more shallow ($f^{-0.42\pm0.02}$).

The loss angle $\varphi_c$ of the high-index material used in each coating,
\begin{equation}
\label{EQ_phiC_CTN}
    \varphi_c(f) = \varphi_c \times \left(\frac{f}{100\textrm{ Hz}}\right)^{\left(1-2\times \textrm{slope}\right)}\ ,
\end{equation}
is extracted from the measurements using the ASD at 100 Hz, the coating structure, a loss angle of 2.3$\times$10$^{-5}$ rad for the low-index material \cite{Granata20}, and an analytic expression for the CTN \cite{Yam15}. Results are reported in Table \ref{TABLE_CTNresults}.
We find, in general, $\varphi_c$ values higher than found with the ring-down method (see Table \ref{TABLEmonolayers}). Although the 100 Hz amplitude is similar among the three samples, there is significantly less of the high-index material in the optimized sample; since we kept the low-index loss angle fixed for our fit, the extracted high-index loss angle increased.

\subsection{Updated Advanced LIGO ETM CTN}
\begin{figure*}
\centering
	\includegraphics[width=1.0\textwidth]{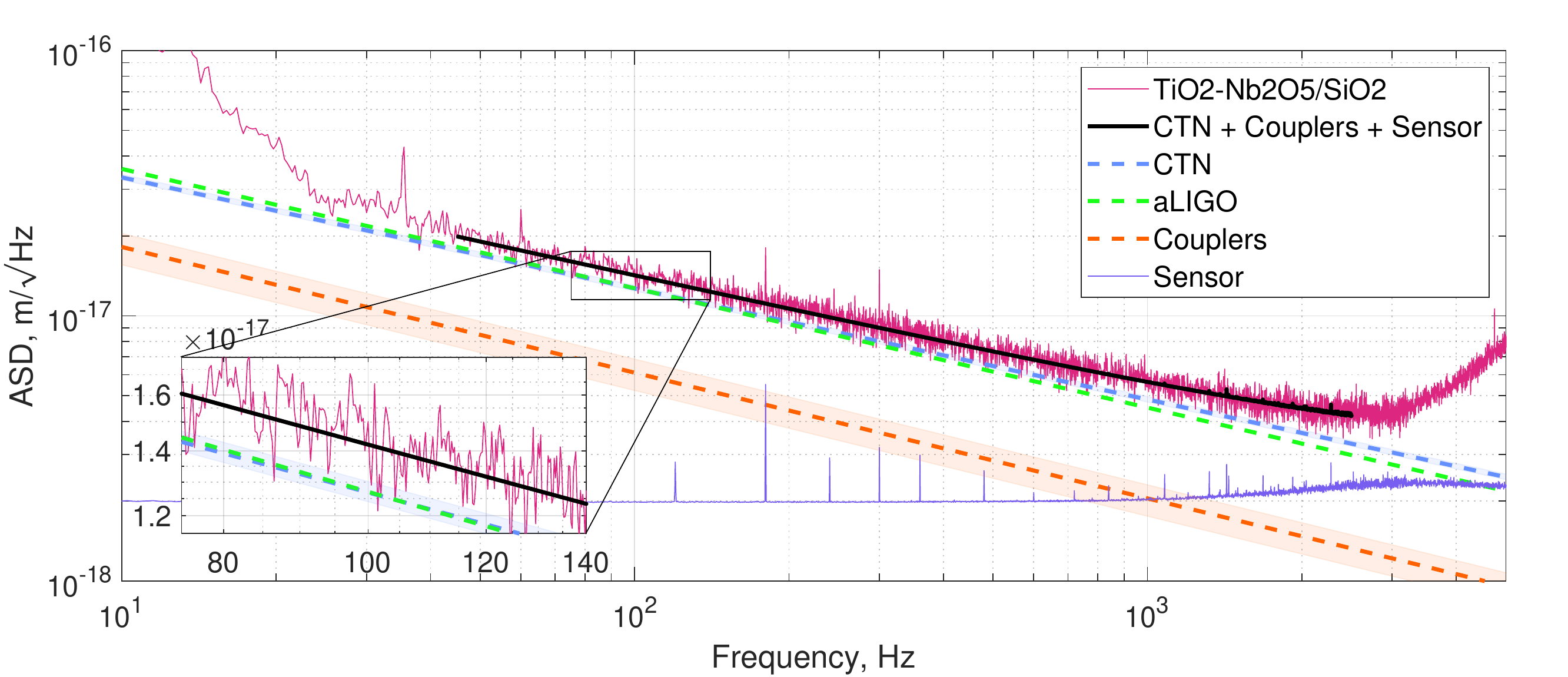}
	\caption{Amplitude spectral density of the TiO$_2$-Nb$_2$O$_5$/SiO$_2$ $\left(\lambda/4\right)$ sample. The sensor noise curve is a combination of noise from electronics and shot noise on the photodetector. The couplers noise curve is the CTN from the cavity end mirrors. The end mirrors contribute CTN that is reduced relative to the sample mirror by a factor of 50. Error bars on the couplers noise curve are 1$\sigma$, and propagate forward to the error on our CTN fit. The black line through the data is the combined CTN + couplers + sensor best fit. The Advanced LIGO curve, updated to properly subtract the coupler contribution, is measured from Advanced LIGO ETM witness samples and presented for comparison. The inset highlights the frequency region around 100 Hz, where CTN most strongly limits the sensitivity of GWDs.}
	\label{FIGctn_asd}
\end{figure*}

Finally, we present an updated value for the frequency-dependent CTN amplitude of the current ETMs used in Advanced LIGO:
\begin{equation}
\label{aLIGOETM}
\left(6.3\pm0.1\right)\times \left( \frac{\text{100 Hz}}{f} \right)^{0.45\pm0.02}\times10{^-21}\frac{\textrm{m}}{\sqrt{\textrm{Hz}}}\ . 
\end{equation}
This is obtained by scaling our measured CTN in Table \ref{TABLE_CTNresults} to the CTN of a 6.2 cm beam on an Advanced LIGO ETM using the methods outlined in in \cite{Gras18}. This new value is slightly lower than the previously reported value \cite{Gras18}, as we are now accounting for the noise contribution of cavity couplers to the measured ASD (see Fig. \ref{FIGctn_asd}).

\section{Discussion}
We developed a set of niobia- and titania-based thin films in order to test in depth their application to the coatings of present and future GWDs, through the measurement of their optical and mechanical properties and thermal noise.

We chose niobia and titania because of their high refractive index, with the aim of minimizing $d$ in Eq.(\ref{eqn.S}). By co-sputtering these two materials, we achieved a 9\% higher refractive index than in tantala-titania layers of current GWDs \cite{Granata20}. Thus, compared to present GWDs, the higher index allowed us to realize thinner HR coatings with comparable optical properties. Eventually, however, all our newly-developed HR coatings unexpectedly showed a very similar CTN level, very close to that of ETMs of Advanced LIGO and Advanced Virgo. This may be explained by the fact that $\varphi_c$ of niobia and titania-niobia layers turned out to be higher than in tantala-titania layers \cite{Granata20}, thus canceling out any improvement derived from having reduced $d$.

There is a significant difference between the loss values of niobia and titania-niobia layers provided by ring-down measurements and direct CTN measurements. Note that the ring-down method used single-layer samples, whereas thermal noise measurement where conducted on HR coatings where many layers are stacked.

The observed discrepancy might come from the silica loss angle used in the analysis of CTN data, which we assumed to be 2.3$\times$10$^{-5}$ rad as measured for silica annealed at 500 $^{\circ}$C \cite{Granata20}. As a reminder, the annealing temperature of our newly-developed HR coatings was limited to 400 $^{\circ}$C because of crystallization. Indeed, as shown in Table \ref{TABLEmonolayers}, the discrepancy between ring-down and CTN values is about 30\% for samples annealed at 400 $^{\circ}$C and about 17\% for samples annealed at 500 $^{\circ}$C. In addition, the discrepancy seems to increase with the total thickness of the low-index material, $d_{\textrm{\tiny{L}}}$, suggesting that the assumed silica loss angle value could be underestimated. However, to make the results match, the silica loss angle would need to be 1.8$\times$10$^{-4}$ rad and 1.1$\times$10$^{-4}$ rad for the samples annealed at 400 $^{\circ}$C and 500 $^{\circ}$C respectively, a factor of 3-8 larger than previously measured \cite{Granata20,Granata16,Penn03}.

An alternative explanation for the observed discrepancy could be the presence of an excess loss in HR stacks, as observed previously on several HR stacks with different designs \cite{Granata16} and especially with the current HR coatings of Advanced LIGO and Advanced Virgo \cite{Granata20}. Although already well-known, the observed excess loss remained unexplained to date and will be the object of further investigation.

In order to be used in future GWDs, $\varphi_c$ of niobia and titania-niobia coatings will have to be reduced. This could be achieved with higher annealing temperatures. For titania-niobia layers, the crystallization might be frustrated either by varying the composition, i.e. the cation ratio Nb/(Ti+Nb), or by adopting a nano-layered structure \cite{Pan14}.

Also, further development will be needed to avoid the presence of defects after annealing, which is a severe obstacle to  implementation in GWDs. Argon trapped in the coating is very likely the cause of such defects \cite{Cevro95,Fazio20}; we are currently working to verify this hypothesis, and to find appropriate solutions.

\section*{Acknowledgments}
This work has been promoted by the Laboratoire des Mat\'{e}riaux Avanc\'{e}s and partially supported by the Virgo Coating Research and Development (VCR\&D) Collaboration. The authors would like to acknowledge the unfailing support and recognition of the LIGO Scientific Collaboration’s optics working group without which this work would not have been possible. The authors also acknowledge the support of the National Science Foundation under awards PHY-1705940 and PHY-0555406. We are also very grateful for the computing support provided by The MathWorks, Inc. This work has document number LIGO-P2000496.

\bibliographystyle{apsrev4-1}

\end{document}